\documentclass[nofootinbib,reprint,amssymb,superscriptaddress,nofootinbib]{revtex4-1}
\usepackage[pdftex]{graphicx}
\usepackage{subfig}
\usepackage{amsmath,amsfonts,amssymb,mathtools}
\usepackage{tasks}
\usepackage{diagbox}
\usepackage{float}
\usepackage{csquotes}
\usepackage{float}
\usepackage{bbm,bm}
\usepackage[normalem]{ulem}
\usepackage{comment}
\usepackage{physics}
\usepackage{xcolor}
\usepackage[utf8]{inputenc}
\usepackage{pgfplots}
\definecolor{blueprl}{RGB}{46,48,146}
\usepgfplotslibrary{groupplots}

\usepackage{lipsum}

\def\ANU{Centre for Quantum Computation and Communication Technology, Department of Quantum Science, Australian National University, Canberra, ACT 2601, Australia.}
  \def\Jena{Institute of Applied Physics, Abbe Center of Photonics, Friedrich-Schiller-Universität Jena, 07745 Jena, Germany}
  
  \def\Fraun{Fraunhofer-Institute for Applied Optics and Precision Engineering IOF, 07745 Jena, Germany}
  \def\MP{Max Planck School of Photonics, 07745 Jena, Germany}
    \def\Astar{Institute of Materials Research and Engineering, Agency for Science Technology and Research (A*STAR), 2 Fusionopolis Way, 08-03 Innovis 138634, Singapore}
\begin{document}
\title{Discriminating mixed qubit states with collective measurements}
\author{Lorc{\'a}n O. Conlon}
\email{lorcanconlon@gmail.com}
\affiliation{\ANU}
\affiliation{\Astar}
\author{Falk Eilenberger}
\affiliation{\Jena}
\affiliation{\Fraun}
\affiliation{\MP}
\author{Ping Koy Lam}
\affiliation{\ANU}
\affiliation{\Astar}
\author{Syed M. Assad}
\email{cqtsma@gmail.com}
\affiliation{\ANU}
\affiliation{\Astar}

\begin{abstract}
It is a central fact in quantum mechanics that non-orthogonal states cannot be distinguished perfectly. This property ensures the security of quantum key distribution. It is therefore an important task in quantum communication to design and implement strategies to optimally distinguish quantum states. In general, when we have access to multiple copies of quantum states the optimal measurement will be a collective measurement. However, to date, collective measurements have not been used to enhance quantum state discrimination. One of the main reasons for this is the fact that, in the usual state discrimination setting with equal prior probabilities, at least three copies of a quantum state are required to be measured collectively to outperform separable measurements. This is very challenging experimentally. In this work, by considering unequal prior probabilities, we propose and experimentally demonstrate a protocol for distinguishing two copies of single qubit states using collective measurements which achieves a lower probability of error than can be achieved by any non-entangling measurement. We implement our measurements on an IBM Q System One device, a superconducting quantum processor. Additionally, we implemented collective measurements on three and four copies of the unknown state and found they performed poorly.
\end{abstract}
\maketitle

\section{Introduction}
\begin{figure*}[t]
\includegraphics[width=\textwidth]{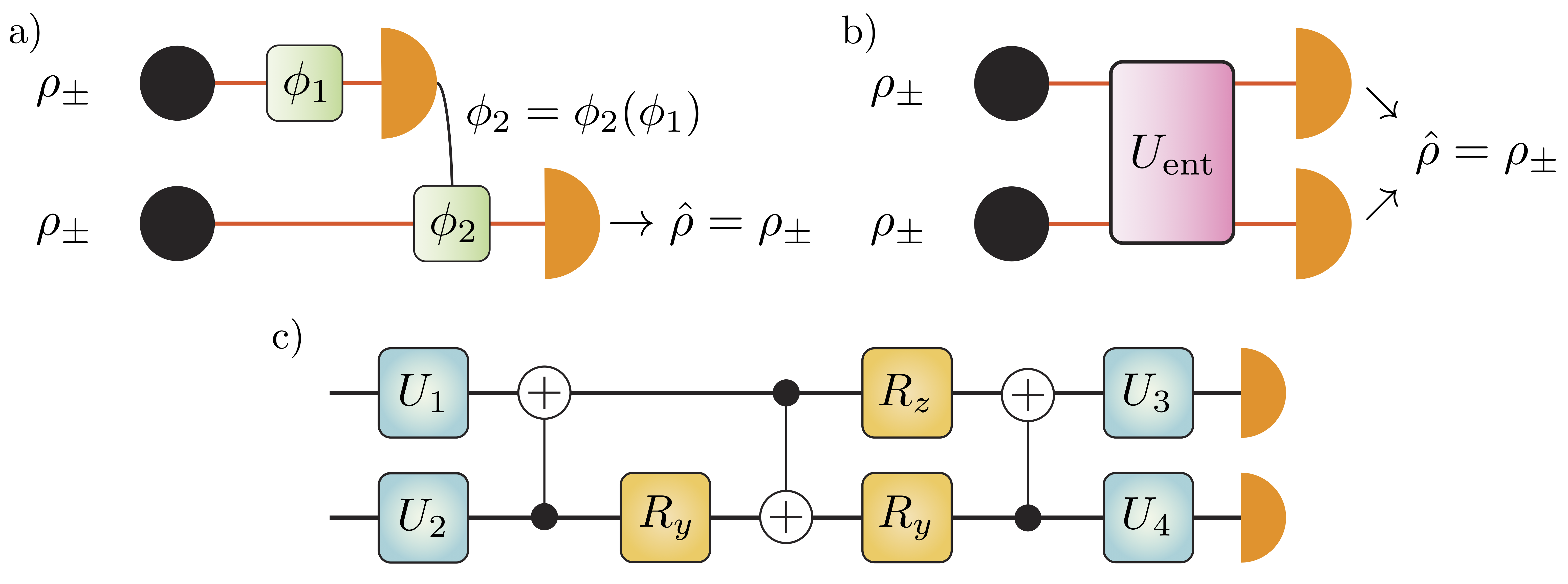}
\caption{\textbf{Schematic for optimal separable and collective measurement state discrimination with two copies.} a) The optimal LOCC measurement when discriminating two copies of a quantum state involves only local operations and classical communication. The measurement angle $\phi_1$ is chosen prior to the experiment to minimise the expected probability of error. Based on the measurement outcome obtained when measuring the first copy, the optimal $\phi_2$ is chosen for the second copy, i.e. $\phi_2=\phi_2(\phi_1)$. A decision of whether the unknown state is $\rho_+$ or $\rho_-$ is made based on the outcome of the second measurement. \mbox{b) The} optimal collective measurement for discriminating two copies of a quantum state requires entangling operations. The decision about what state is present is made based on all measurement results. c) Quantum circuit used to implement the optimal collective measurements shown in b). The $U$ gates correspond to arbitrary single qubit unitary matrices and the $R_y$ and $R_z$ gates correspond to single qubit rotations. The parameters of the circuit are numerically optimised to implement the optimal POVM for a given set of parameters $(q,\alpha,v)$.}
\label{fig:scen1}
\end{figure*}

Quantum state discrimination, or hypothesis testing, was first introduced by Helstrom~\cite{helstrom1969quantum}. It can be described in terms of a simple two party game. Alice sends $M$ copies of one of two states, $\rho_+$ or $\rho_-$, to Bob, with some specified prior probability. Bob then tries to implement the optimal measurement to decide which state he has received according to some figure of merit. Owing to the no-cloning theorem~\cite{wootters1982single} if the two states to be distinguished are not orthogonal, they can't be distinguished perfectly. This means that there will necessarily be some error when Bob decides which state he has received. Typically, Bob tries to minimise the average probability of error. When distinguishing between two pure orthogonal states, it is known that a single separable measurement, consisting of only local operations and classical communication (LOCC), can distinguish between the two states perfectly~\cite{walgate2000local}. \footnote{Note that while LOCC usually refers to operations and communication between two parties Alice and Bob, in this work local operations means operations on a single mode of Bob's state and classical communication means Bob cannot implement entangling operations between the different modes of the state he receives.}

The situation becomes considerably more interesting when considering non-orthogonal states. For distinguishing multiple pure non-orthogonal states, it was first conjectured by Peres and Wooters that collective measurements can improve the error probability compared to LOCC measurements~\cite{peres1991optimal}. A collective measurement here refers to an entangling measurement acting on $M$ copies of the state, $\rho_+^{\otimes M}$ or $\rho_-^{\otimes M}$, simultaneously. In contrast, a LOCC measurement means that the $M$ copies of the state received by Bob are measured individually, and each subsequent measurement may be adaptively updated based on previous measurement results. Hence, collective measurements are more general than LOCC measurements. For distinguishing a continuous spectrum of different quantum states given a finite number of copies, the advantage of collective measurements over separable measurements was confirmed in Ref.~\cite{massar1995optimal}. However, for distinguishing two non-orthogonal pure states, it is known that collective measurements offer no advantage over separable measurements~\cite{brody1996minimum,ban1997accessible,acin2005multiple}.

When we consider distinguishing between mixed states, the situation becomes more interesting again. For distinguishing two mixed states, collective measurements generally offer an advantage over separable measurements. Despite theoretical progress on the role of collective measurements for discriminating between mixed states~\cite{calsamiglia2008quantum,calsamiglia2010local,higgins2011multiple,flatt2019multiple}, experimental progress along this avenue has been limited. A great deal of experimental work has focused on optimally distinguishing single copies of pure states~\cite{cook2007optical,wittmann2008demonstration,bartuuvskova2008programmable,waldherr2012distinguishing,becerra2013implementation,izumi2020experimental,izumi2021adaptive,sidhu2021quantum,gomez2022experimental} or using adaptive separable measurements to distinguish multiple copies of either pure~\cite{becerra2013experimental} or mixed~\cite{higgins2009mixed} states. The multi-copy discrimination of coherent and thermal states was considered in Ref~\cite{jagannathan2022demonstration}. However, in this case, the multiple copies were measured individually, not with a collective measurement and the multi-copy part refers only to the decision making process.

Thus, all previous state discrimination experiments have been limited to LOCC measurements which may not attain the ultimate limits for distinguishing non-orthogonal mixed states. A major reason for this is that collective measurements are difficult to implement experimentally. Indeed, although collective measurements are known to aid a range of tasks in quantum information, including quantum metrology~\cite{conlon2021efficient}, Bell violations~\cite{liang2006better} and entanglement distillation~\cite{bennett1996purification}, they have only recently been experimentally demonstrated~\cite{roccia2017entangling,hou2018deterministic,conlon2023approaching}.\footnote{There are other tasks which benefit from measurements in an entangled basis, not necessarily on multiple copies of the same quantum state, including some quantum metrology problems~\cite{marciniak2022optimal}, quantum orienteering~\cite{gisin1999spin,jeffrey2006optical,tang2020experimental} and quantum communications~\cite{delaney2022demonstration,crossman2023quantum}.} This advantage is thus far restricted to performing collective measurements on only two copies of a quantum state~\cite{conlon2023approaching}.  Previous investigations into the use of collective measurements for state discrimination suggest that collective measurements only offer an advantage over LOCC measurements when measuring at least three copies of the quantum state simultaneously~\cite{higgins2009mixed,calsamiglia2010local,higgins2011multiple}. However, as we shall show, this is only true when the prior probabilities for the unknown state being either $\rho_+$ or $\rho_-$ are equal.

 

In this work, we bypass these difficulties and experimentally demonstrate state discrimination enhanced by collective measurements. That is, we implement a collective measurement which distinguishes between two copies of a qubit state with some known prior probability, either $\rho_+^{\otimes 2}$ or $\rho_-^{\otimes 2}$, with an error probability lower than what is possible by any LOCC measurement on the same states. A conceptual schematic of both two-copy LOCC and two-copy collective measurements is shown in Fig.~\ref{fig:scen1}. To realise our goal, we first find a regime where two copy collective measurements are able to outperform any LOCC measurement. This is done through the dynamic programming approach introduced in Ref.~\cite{higgins2009mixed}. We then convert the optimal collective measurements to quantum circuits using the same techniques as described in Refs.~\cite{conlon2023approaching,vatan2004optimal}. The same process is followed when implementing three- and four-collective measurements. Finally, these quantum circuits are implemented on the Fraunhofer IBM Q System One (\mbox{F-IBM QS1}) processor. This is in line with a recent trend of using such processors to test otherwise hard to reach aspects of quantum physics, such as quantum metrology~\cite{conlon2023approaching,li2023optimal}, quantum foundations~\cite{alsina2016experimental,ku2020experimental,sadana2022testing}, quantum chemistry~\cite{kandala2017hardware,kandala2019error}, quantum optics~\cite{cholsuk2023efficient} and quantum network simulations~\cite{baumer2021demonstrating}. 



%
\begin{figure*}[t]
\includegraphics[width=0.8\textwidth]{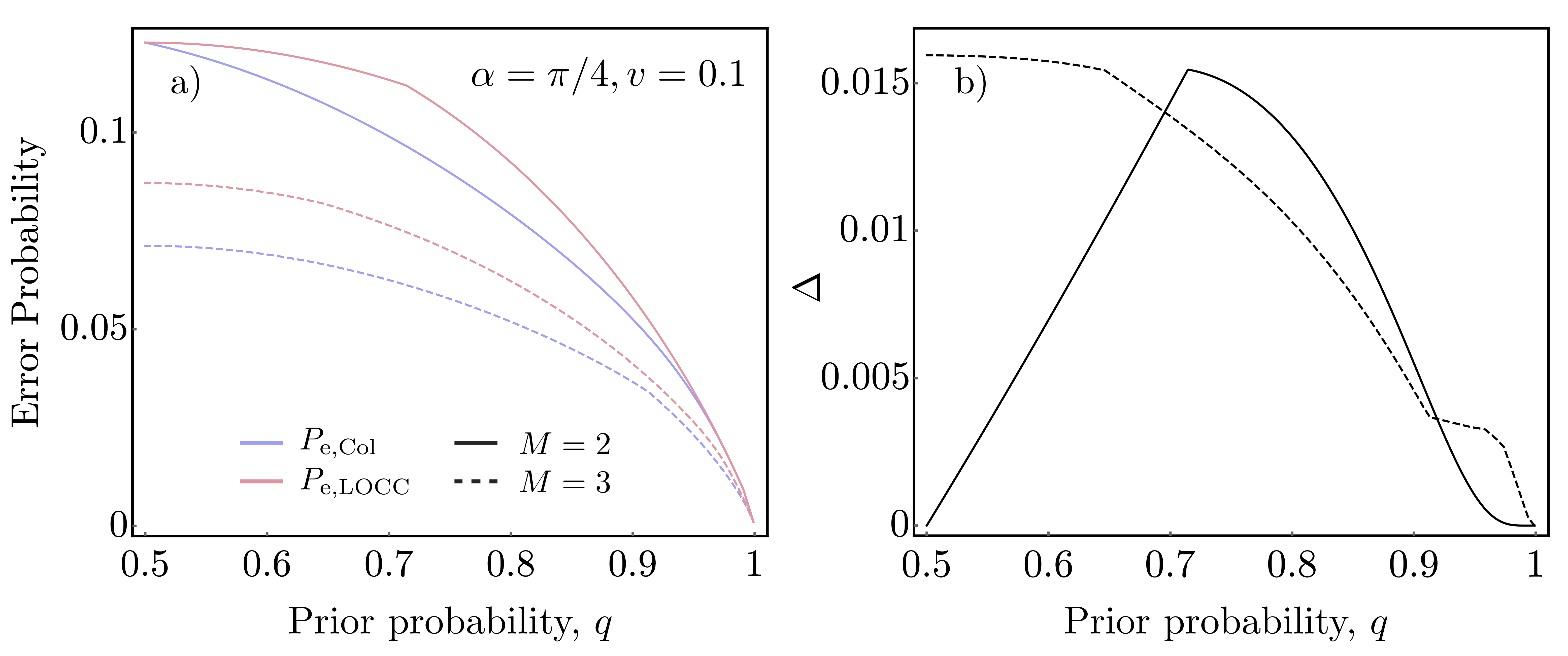}
\caption{\textbf{Experimentally feasible regime for demonstrating state discrimination enhanced by collective measurements.} a) The expected error probabilities attainable for distinguishing between two and three copies of the quantum state, $\rho_\pm^{\otimes 2}$ and $\rho_\pm^{\otimes 3}$, when allowing only LOCC measurements and when allowing collective measurements. b) We plot the gap between the optimal LOCC measurement and the optimal collective measurement for two and three copies of the probe state. These plots correspond to $\alpha=\pi/4$ and $v=0.1$. In both figures the solid and dashed lines correspond to two- and three-copy results respectively.}
\label{fig:delta}
\end{figure*}
\section{Theoretical results}
\subsection{Theoretical model}
We follow the model of Higgins\textit{ et al}~\cite{higgins2011multiple}, where we wish to discriminate between the two qubit states described by
\begin{equation}
\rho_{\pm}=\frac{1}{2}\bigg(\mathbb{I}_2+(1-v)(\sigma_3\text{cos}(\alpha)\pm\sigma_1\text{sin}(\alpha))\bigg)\;,
\end{equation}
where $\mathbb{I}_2$ is the $2\times2$ identity matrix, $\sigma_i$ denotes the $i$th Pauli matrix, and $v$ and $\alpha$ describe the extent to which the state is mixed and the angle between the two states in Hilbert space respectively. Note that $v=0$ corresponds to a pure state and $v=1$ represents the maximally mixed state. Given $M$ copies of the above state, $\rho_{\pm}^{\otimes M}$, our task is to decide which state we are given with the minimum probability of error. Denoting the prior probability for the unknown state to be the state $\rho_{+}$ as $q$, the error probability is given by
\begin{equation}
\label{eq:errorprobabilitybasic}
P_\text{e}=qP_{-|+}+(1-q)P_{+|-}\;,
\end{equation}
where $P_{-|+}$ is the probability of guessing the state $\rho_{-}$ when given the state $\rho_{+}$ and $P_{+|-}$ is similarly defined.

\subsection{Optimal collective measurement}
In general, the optimal measurement to distinguish $\rho_{\pm}^{\otimes M}$, will be a collective measurement which involves entangling operations between all $M$ modes. The minimum error probability, in this case, is given by the Helstrom bound~\cite{helstrom1969quantum} 
\begin{equation}
\label{eq:helstromdefinition}
P_\text{e,Col}=\frac{1}{2}\bigg(1-\left\lVert q\rho_+^{\otimes M}-(1-q)\rho_-^{\otimes M}\right\rVert\bigg)\;,
\end{equation}
where $P_\text{e,Col}$ denotes the error probability which can be attained by collective measurements and $\lVert A\rVert$ is the sum of the absolute values of the eigenvalues of $A$. It is known that the Helstrom bound can be saturated by measuring the following observable, which may or may not correspond to a collective measurement
\begin{equation}
\label{eq:helstromobservable}
\Gamma=q\rho_{+}^{\otimes M}-(1-q)\rho_-^{\otimes M}\;.
\end{equation}
The state $\rho_{\pm}$ is guessed according to the sign of the measurement output. Note that, in some scenarios, $\Gamma$ will be either positive or negative meaning that the best strategy is always simply to guess the corresponding state. In practice, we implement the operator $\Gamma$, through a positive operator valued measure (POVM). A POVM is a set of positive operators $\{\Pi_k\geq0\}$, which sum to the identity $\sum_k\Pi_k=\mathbb{I}$. In general, to experimentally implement an arbitrary POVM, Naimark's theorem~\cite{neumark1943spectral} is used to convert the POVM to a projective measurement in a higher dimensional Hilbert space. However, it transpires that for the purposes of this work, the number of POVM elements necessary to saturate Helstrom's bound is less than or equal to the dimensions of the Hilbert space and so Naimark's theorem is not needed. Using the optimal measurement, given in Eq.~\eqref{eq:helstromobservable}, and the definition of error probability in Eq.~\eqref{eq:errorprobabilitybasic}, we can verify that the Helstrom bound becomes $(1-\lVert\Gamma\rVert)/2$, as expected from Eq.~\eqref{eq:helstromdefinition}~\cite{bergou2010discrimination}. In general the POVM required to implement $\Gamma$ will correspond to a collective measurement. In this work we implement two-copy collective measurements, as depicted in Fig~\ref{fig:scen1} b).


\subsection{Optimal separable measurement}
A separable measurement is any measurement where entangling operations are not allowed, i.e. only LOCC are allowed. Although there are many different types of LOCC measurements~\cite{higgins2011multiple}, for the purposes of this work, we need only consider the optimal separable measurement. In general, this requires measurements which are optimised depending on the number of copies available. When finding the optimal LOCC measurement for a large number of copies, it can be beneficial to use the dynamic programming approach from Ref.~\cite{higgins2009mixed}. As we are concerned with LOCC measurements on only a small number of copies of the state it is easy to verify the optimal measurement by a brute force computation. A description of the simulation approaches used to find the optimal LOCC measurements is provided in appendix~\ref{apen:numappr}.

For the scenario we are considering, the individual measurements are parametrised by a single angle. We need only consider the measurements defined by the projectors $\Pi_0=\ket{\psi}\bra{\psi}$ and $\Pi_1=\ket{\psi^\perp}\bra{\psi^\perp}$ where
\begin{equation}
\label{eq:meass}
\ket{\psi}=\text{cos}(\phi)\ket{0}+\text{sin}(\phi)\ket{1}\;,
\end{equation} 
and $\bra{\psi}\ket{\psi^\perp}=0$. For two-copy LOCC measurements the optimal angle used in the second measurement will depend on the measurement outcome of the first POVM as shown in Fig~\ref{fig:scen1} a). Either $\Pi_0$ or $\Pi_1$ will click in the first measurement, giving two different posterior probabilities, which will each have their own optimal measurement angle. After the second measurement the posterior probabilities are updated again and the more likely state is chosen as our guess. Therefore, the optimal two-copy LOCC measurement is described by three measurement angles, which can be numerically optimised to find the minimum expected LOCC error probability. 

\subsection{Gap between collective and separable measurements}
Knowledge of the optimal separable and collective two-copy measurements enables us to find an experimentally feasible regime for demonstrating the advantage of collective measurements. This is done by operating in the regime of unequal priors, i.e. $q\neq0.5$. This operating regime was found through the rather surprising observation that at $q=0.5$, for two copies, LOCC measurements perform the same as collective measurements but for three copies collective measurements outperform LOCC measurements. The expected error probabilities for both collective and LOCC measurements for distinguishing two copies and three copies of the quantum state are shown in Fig.~\ref{fig:delta} a). To investigate this behaviour further, we define the gap between LOCC measurements and collective measurements for $M$ copies of the quantum state (in terms of minimum attainable error probability), as
\begin{equation}
\label{eq:delta}
\Delta_M=P_{\text{e,LOCC}}(M)-P_{\text{e,Col}}(M)\;,
\end{equation}
where $P_{\text{e,LOCC}}(M)$ corresponds to the minimum error probability which can be attained with LOCC measurements given $M$ copies of the unknown state, $\rho_{\pm}^{\otimes M}$. We plot this quantity in Fig.~\ref{fig:delta} b) for a particular value of $\alpha$ and $v$. Further plots of $\Delta$ for different sets of $\alpha$ and $v$ values are presented in appendix~\ref{apen:delta}. The optimal LOCC measurement is obtained using dynamic programming, with 2501 entries in the measurement angle and error probability tables. The number of entries in these tables corresponds to how discretised the space of possible measurement angles is. A greater number of entries therefore corresponds to greater numerical accuracy. 2501 entries was found to be a sufficiently large number of entries to ensure the optimality of our solutions to 9 significant figures. While, the differences between two and three-copy measurements in Fig.~\ref{fig:delta} is interesting in and of itself, the key take-away here is that two-copy collective measurements can offer an advantage over LOCC measurements when $q\neq0.5$. This is important as two-copy collective measurements are at the limit of what is experimentally feasible~\cite{conlon2023approaching}.

\section{Experimental Results}
\subsection{Two-copy collective measurement results}
It is always possible to saturate the Helstrom bound by implementing the observable given in Eq.~\eqref{eq:helstromobservable}. However, mapping this observable to an experimental set-up is not straightforward. To circumvent this, we numerically find a projective measurement that saturates the Helstrom bound. From this, the unitary matrix we need to implement can be found in a simple manner, see e.g. Ref.~\cite{conlon2023approaching}. Finally, we convert the necessary unitary matrix to experimentally implementable quantum circuits using the decomposition of Ref.~\cite{vatan2004optimal}. This involves finding 15 free parameters and requires only three CNOT gates as is shown in \mbox{Fig.~\ref{fig:scen1} c)}. We then implement these circuits on the \mbox{F-IBM QS1} processor. When measuring in the computational basis there are four possible measurement outcomes, $\ket{00},\ket{01},\ket{10}$ and $\ket{11}$. 

\begin{figure*}[t]
\includegraphics[width=\textwidth]{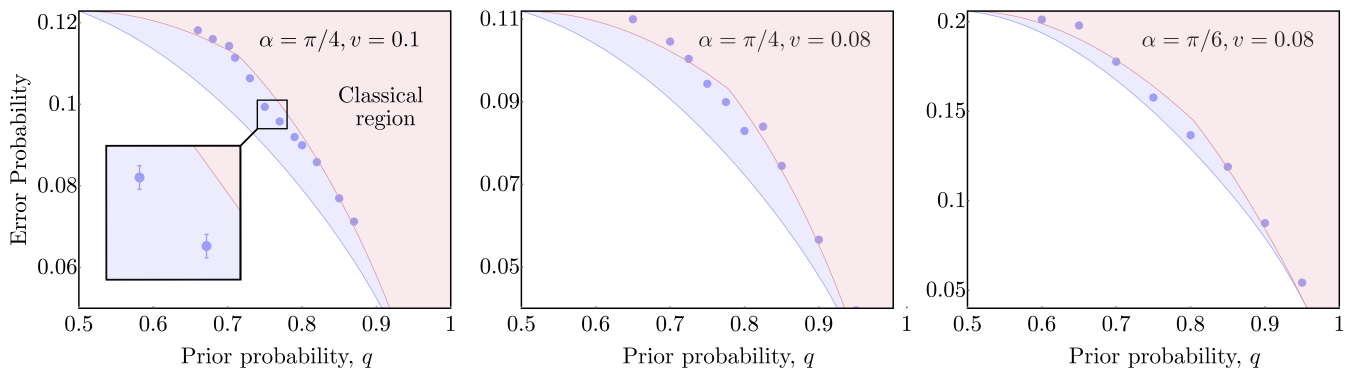}
\caption{\textbf{Experimental data for discriminating quantum states using two-copy collective measurements.} In all three figures the shaded red region corresponds to the error probabilities accessible by LOCC measurements and the blue region corresponds to the error probabilities attainable by collective measurement. The blue markers correspond to the experimental data. Each data point corresponds to 200,000 shots. All data points include statistical error bars corresponding to one standard deviation obtained via bootstrapping, but in most cases the data points are larger than the error bars. The inset is included in the first figure to show the scale of the error bars. Each data point corresponds to a POVM optimised for that particular prior probability. }
\label{fig:data}
\end{figure*}

We assign the outcome $\ket{00}$ to the state $\rho_-$ and the remaining outcomes are assigned to the state $\rho_+$, i.e. if the circuit output is $\ket{00}$ we guess that the unknown state was $\rho_-$, for all other measurement results we guess $\rho_+$. By preparing the states $\rho_+$ and $\rho_-$ many times, the average error probability can be extracted. We repeat this for several different prior probabilities using an optimised POVM for each prior. For each prior probability, we use 200,000 shots, to ensure we obtain an accurate estimate of the average error probability. The results of these experiments are shown in Fig.~\ref{fig:data} for three different sets of $\alpha$ and $v$ values. The red shaded regions in Fig.~\ref{fig:data} correspond to the minimal error probability which can be achieved by any LOCC measurement. The blue shaded regions correspond to the error probability which can be attained by the optimal collective measurements. In theory, our quantum circuits should be able to reach the minimum error probability set by the blue line. However, as can be seen in Fig.~\ref{fig:data}, our experiment does not reach the theoretical collective measurement limit. This is due to a combination of gate errors and readout error in the F-IBM QS1 device. Nevertheless, in several instances our collective measurements surpass what is possible by LOCC measurements. For example, the two data points shown in the inset of Fig.~\ref{fig:data} surpass the expected error probability attainable by LOCC measurements by more than six standard deviations. To the best of our knowledge, this is the first time this has been achieved for quantum state discrimination.

However, we note that not all of our experiments were successful in showing the advantage of collective measurements. For some values of $\alpha$ and $v$, our collective measurements achieved an error probability which was comparable to the best LOCC measurement. These data points are included in appendix~\ref{apen:extres}.
 \begin{figure}[tp!]
\includegraphics[width=0.45\textwidth]{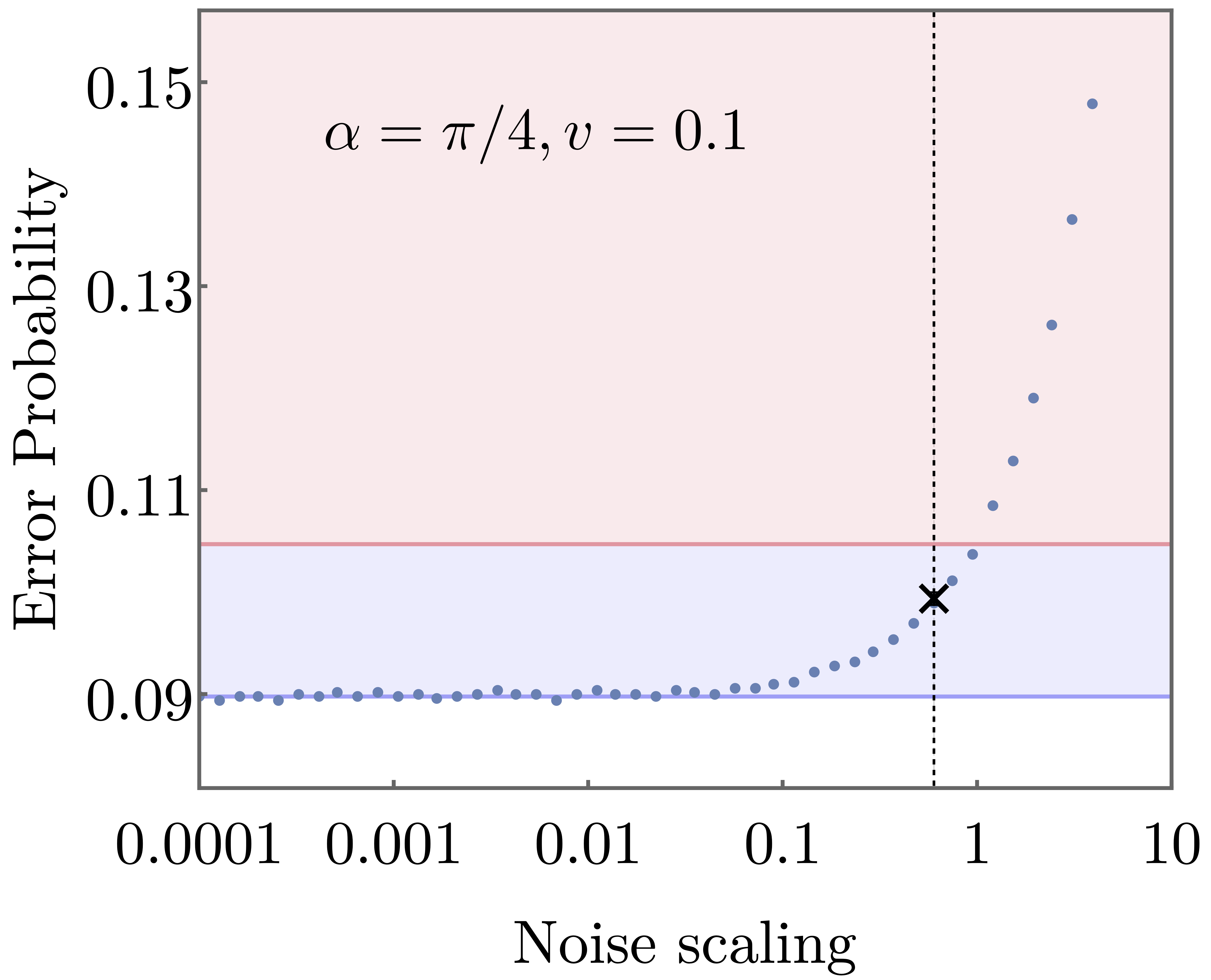}
\caption{\textbf{Simulation of two-copy collective measurement with noise.} The error probability is shown as a function of the noise scaling term, with the default noise parameters, corresponding to a scaling of 1, specified in the main text. The two states being discriminated are characterised by $v=0.1$, $\alpha=\pi/4$ and $q=0.75$. The shaded red and blue regions correspond to the limits of LOCC and collective measurements respectively. The black cross corresponds to the experimentally obtained error probability, and the error bars are smaller than the marker size. The vertical positioning of the experimental data points is of no physical significance. The dashed black vertical line corresponds to a noise scaling of 0.6. }
\label{fig:apend:noisysim1}
\end{figure}
\subsection{Simulation of two-copy collective measurement with noise}
\label{apen:noisysim1}
We now present a noisy simulation of one particular implementation of our two-copy collective measurement to provide an indication of the noise level needed to reach the Helstrom bound in future experiments. We use \mbox{IBM Q's} QISKIT package to model the noise, which includes readout error and depolarising noise on all gates. We choose a starting noise level that approximates the device used in our experiments: single qubit gate error rates of $10^{-3}$, two qubit gate error rates of $10^{-2}$, $\ket{0}$ state readout error of $10^{-2}$ and $\ket{1}$ state readout error of $2\times10^{-2}$. With these parameters, we model the error probability which can be attained for $v=0.1$, $\alpha=\pi/4$ and $q=0.75$. The simulation was then repeated, scaling all the error rates by a noise scaling term, as shown in Fig.~\ref{fig:apend:noisysim1}. As expected, when the noise is scaled down sufficiently, the two-copy collective measurement approaches the Helstrom bound. These simulations suggest that an improvement in error rates by approximately one order of magnitude would allow the two-copy Helstrom bound to be saturated. However, this comes with the caveat that the noise model used may be missing important features. 

\begin{figure}[t!]
\includegraphics[width=0.47\textwidth]{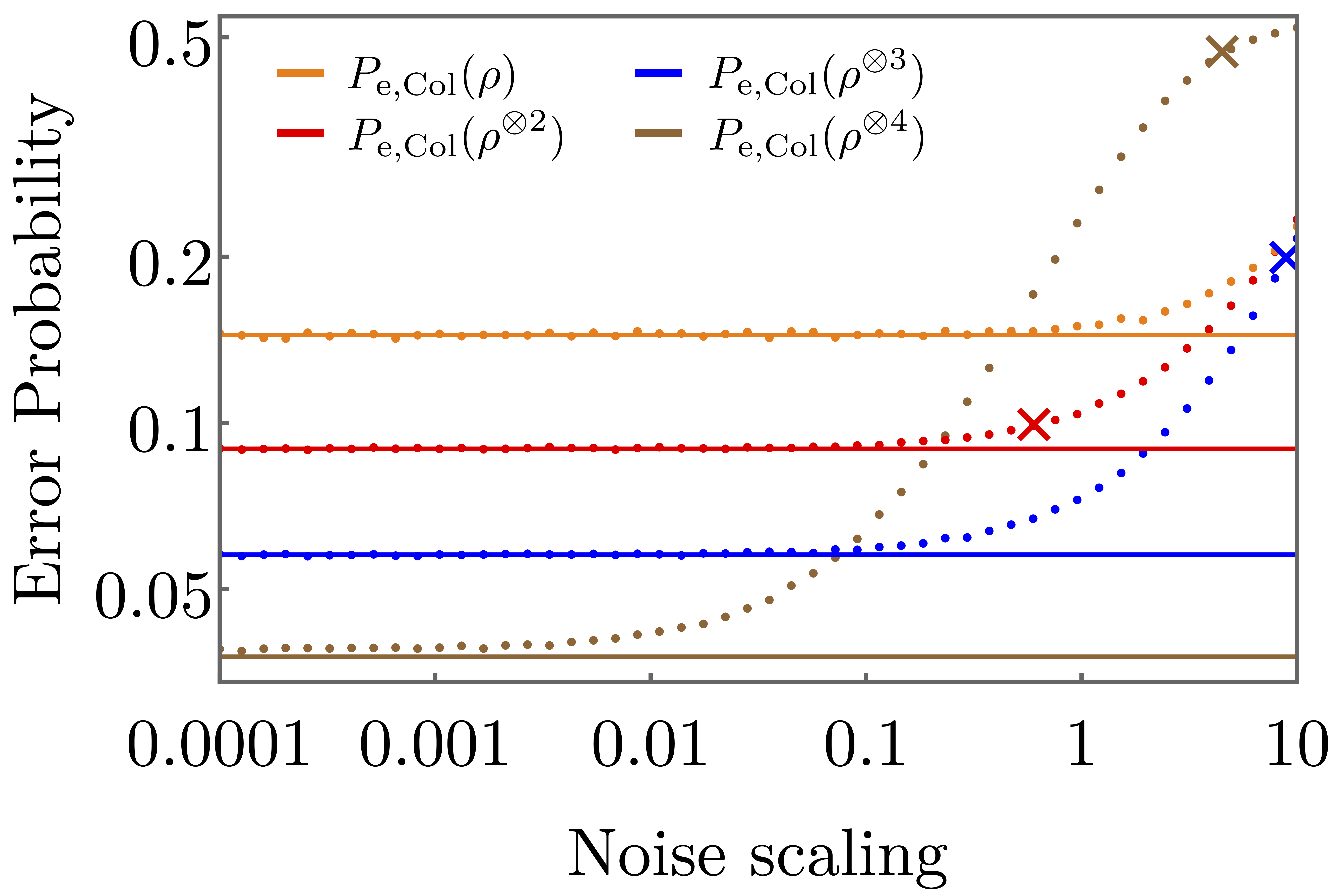}
\caption{\textbf{Three- and four-copy collective measurements - experiments and simulation.} The error probability is shown as a function of the noise scaling term, with the default noise parameters, corresponding to a scaling of 1, specified in the main text. The two states being discriminated are characterised by $v=0.1$, $\alpha=\pi/4$ and $q=0.75$. The crosses correspond to the experimentally obtained error probability and all error bars are smaller than the marker size. The vertical positioning of the experimental data points is of no physical significance. The solid horizontal lines show the theoretical limit given by the Helstrom bound for the corresponding number of copies.  }
\label{fig:apend:noisysim2}
\end{figure}

\subsection{Three- and four-copy collective measurements}
\label{apen:ncopynoisysim}
Finally, we extend the above results to three- and four-copy collective measurements. The three- and four-copy measurements that we implement were found by searching for projective measurements in a three and four qubit Hilbert space respectively. In the ideal case, the three-copy measurement found is able to saturate the three-copy Helstrom bound. The four-copy measurement cannot saturate the four-copy Helstrom bound, however, it can surpass the optimal four-copy LOCC measurement. To saturate the four-copy Helstrom bound with projective measurements would likely require ancilla qubits, which would significantly increase the complexity of the corresponding quantum circuit. For $v=0.1$, $\alpha=\pi/4$ and $q=0.75$, we implemented these three- and four-copy measurements on the \mbox{F-IBM QS1} device, with the results presented in Fig.~\ref{fig:apend:noisysim2}. Notably, both the three- and four-copy measurement perform worse than measuring only a single qubit. This suggests that when given many copies of a state, rather than trying to implement a very complex entangling measurement, it is better to measure the quantum states in smaller groups. Even in an optimistic scenario, we will need an order of magnitude reduction in noise levels for the four-copy entangling measurement to be useful.\footnote{If it was known in advance that our measurement device is extremely noisy, we can always attain an error probability of 0.25 as, in this case, we can simply ignore our measurement results and guess the more likely state. }

In contrast to the three-copy experimental results, the simulation of the three-copy measurement, using the same noise parameters as the previous section, performed quite well. This highlights the limitations of our noise model, particularly for circuits involving larger numbers of qubits. For example, qubit connectivity becomes an important factor in practical three-copy measurements. There is also a significant gap between the experimental results and the noisy simulation in the four-copy case. These experiments and simulations suggest that while it is possible to surpass the LOCC limits using two-copy collective measurements, the ultimate limits in state discrimination shall remain out of reach for the foreseeable future.

%

\section{Conclusion}
Recent experimental advances have led to collective measurements offering an advantage in several areas of quantum information~\cite{roccia2017entangling,hou2018deterministic,conlon2023approaching,parniak2018beating,wu2019experimentally,yuan2020direct,wu2020minimizing}. However, prior to this work, this advantage had not been extended to quantum state discrimination. By utilising the \mbox{F-IBM QS1} superconducting processor and operating in the regime of unequal prior probabilities, we have been able to experimentally distinguish two copies of two quantum states with an error probability smaller than the best possible LOCC measurement. The further development of this capability may aid numerous applications of quantum state discrimination, including quantum illumination~\cite{lloyd2008enhanced,tan2008quantum,bradshaw2021optimal,pirandola2018advances} and exoplanet detection~\cite{huang2021quantum}.

There are several natural avenues for extending this work. In this work, we have taken, as our figure of merit, the average error in distinguishing two states, but several other figures of merit are commonly used~\cite{barnett2009quantum,bae2015quantum}, including unambiguous state discrimination~\cite{ivanovic1987differentiate,dieks1988overlap}, asymmetric state discrimination~\cite{karsa2020quantum} and, state discrimination using the minimum number of copies~\cite{slussarenko2017quantum}. When decomposing the three- and four-copy measurement circuits, we simply used the default decomposition provided on QISKIT. It may be possible to obtain an approximate circuit with a similar theoretical minimum error probability, but requiring a greatly reduced number of CNOT gates~\cite{gundlapalli2022deterministic}. Furthermore, this work may be extended to examine the role of collective measurements when distinguishing more than two states~\cite{andersson2002minimum} or for distinguishing states which live in a larger Hilbert space, as was done for two qubit states in Ref.~\cite{patterson2021quantum}. In this work, we have experimentally investigated the role of collective measurements for only a small number of copies of the quantum state. However, going beyond the few-copy setting and investigating the asymptotic attainability of the ultimate limits in quantum state discrimination, as has recently been done in quantum metrology~\cite{conlon2022gap}, may reveal other important features of state discrimination.

Note added: While this manuscript was under review we became aware of Ref.~\cite{tian2023minimum}. In this paper two-copy collective measurements were used for minimum consumption state discrimination.

\section*{Data availability}
Any data obtained for this project are available from the corresponding authors upon request.
\section*{Code availability}
Any code used for this project are available from the corresponding authors upon request.
\section*{Acknowledgements}
We acknowledge the use of IBM Quantum services for this work. The views expressed are those of the authors, and do not reflect the official policy or position of IBM or the IBM Quantum team.

This research was funded by the Australian Research Council Centre of Excellence CE170100012, Laureate Fellowship FL150100019 and the Australian Government Research Training Program Scholarship.

This work has been supported by the Fraunhofer-Gesellschaft zur Förderung der angewandten Forschung e.V via the QuantumNow programme and the German Federal Ministry of Education and Research, FKZ: 13XP5053A. This work was funded by the Deutsche Forschungsgemeinschaft (DFG, German Research Foundation) as part of CRC 1375 NOA.

\appendix
\section{Details of numerical simulations for finding the optimal LOCC measurement}
\label{apen:numappr}
In this appendix we give a detailed description of how the expected error probabilities attainable by LOCC measurements are computed. We first describe the dynamic programming approach of Refs.~\cite{higgins2009mixed,calsamiglia2010local,higgins2011multiple}. This technique allows us to break the overall optimisation problem into smaller optimisation problems which can be solved in turn. We wish to find the measurement angles which minimise the expected probability of error when allowing LOCC measurements on multiple copies of the state $\rho$.

We define $q_i$ as the probability that the unknown state is the state $\rho_{+}$ just before performing the $i$th measurement. Hence, the initial prior probability that the unknown state is $\rho_{+}$, which is denoted $q$ in the main text, is denoted $q_1$ in this appendix. After performing each measurement, the posterior probability of the unknown state being $\rho_{+}$ will change, depending on the measurement result. At each stage, we wish to find the optimal angle $\phi$ for the measurement given by Eq.~\eqref{eq:meass}. The optimal angle for the $i$th measurement, $\phi_i$, is that which minimises the average error probability and will depend on the number of copies available $M$ and the prior probability $q_{i}$. For every measurement there are two possible outcomes, $\Pi_0(\phi)=\ket{\psi}\bra{\psi}$ or $\Pi_1(\phi)=\ket{\psi^\perp}\bra{\psi^\perp}$, where we have made the dependence on $\phi$ explicit. The probability of each outcome is given by $\text{tr}\{\rho\Pi_i\}$. We shall denote the outcome of the $i$th measurement as $D_i$, so that $D_i$ can correspond to either $\Pi_0$ or $\Pi_1$ clicking. Given the measurement outcome $D_i$, the posterior probability after performing the $i$th measurement (or equivalently the prior probability for the following measurement) is given by
\begin{equation}
q_{i+1}=\frac{\text{Pr}[D_i|\rho_{+},\phi_i]q_{i}}{\text{Pr}[D_i|q_{i},\phi_i]}\;,
\end{equation}
where $\text{Pr}[D_i|a]$ is the probability of observing the measurement outcome $D_i$ given the conditions $a$.

%

Given $M$ total states to measure, we can measure $m$ ($m\leq M$) states to obtain an intermediate error probability. We denote the expected error probability (with $M-m$ states available to measure) as $R_{m}$. Using this notation, $R_M$ is the error probability when there are no states left to measure. At this point, the optimal strategy is simply to guess whichever state is more likely, so that
\begin{equation}
\label{eq:min}
R_M=\text{min}(q_{M+1},1-q_{M+1})\;.
\end{equation}
Note that at each measurement stage $R_m$ depends on $q_{m+1}$, which is determined by the preceding measurement angles $\phi_1, \phi_2(D1), \phi_3(D1,D2) \hdots \phi_M(D1,D2,..,D_{M-1})$. The expected error probability at each previous measurement step can be calculated from the error probability at the current step
\begin{equation}
\label{eq:R_recursive}
R_{m-1}(q_m)=\sum_{D_m}\text{Pr}[D_m|q_m,\phi_m]R_m(q_{m+1})\;.
\end{equation}
Our aim is to minimise the expected error probability when all $M$ copies of the quantum state are available to measure, denoted $R_0$. If we have only a single copy of the quantum state left to measure, the optimal measurement angle is known to be~\cite{higgins2011multiple}
\begin{equation}
\phi_{M,\text{opt}}(q_M)=\frac{1}{2}\text{arccot}((2q_M-1)\text{cot}(\alpha))\;.
\end{equation}
This allows $R_{M-1}$ to be calculated. $R_{M-2}$ can then be calculated by minimising Eq.~\eqref{eq:R_recursive}. Each measurement stage can be recursively optimised in this manner. Continuing in this manner gives $R_0(q_1)$ as the expected error probability. 

\begin{figure}[t!]
\includegraphics[width=0.5\textwidth]{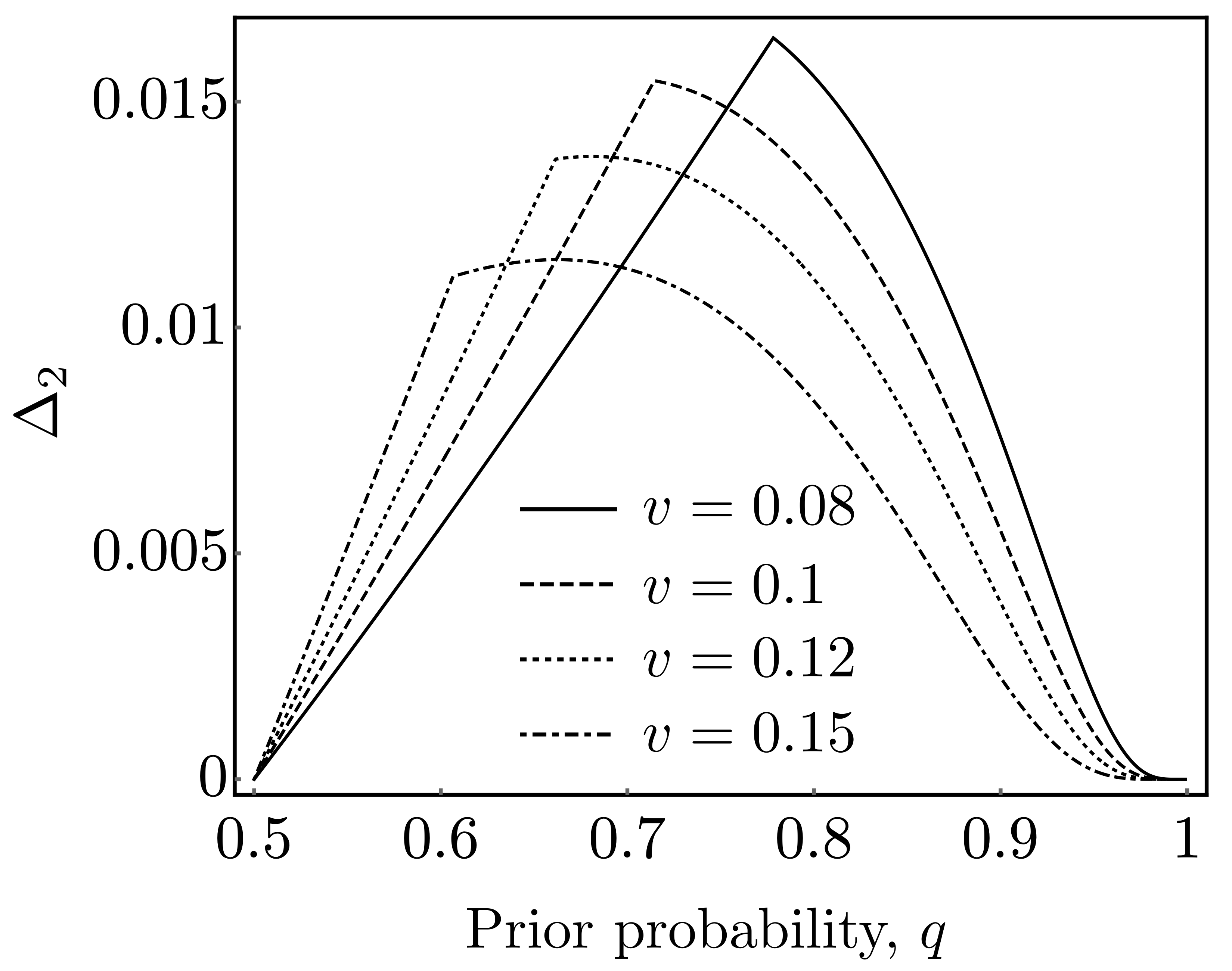}
\caption{\textbf{Gap between the optimal LOCC measurement and the optimal collective measurement for two copies of the probe state.} All figures correspond to $\alpha=\pi/4$, and the $v$ value for each line is indicated in the legend.}
\label{fig:apendelta}
\end{figure}

Dynamic programming offers an efficient way to compute the optimal angles. In the two-copy case it is possible to do a brute force search for the optimal angles to verify the correctness of our simulations. A simple expression for the expected error probability in the two-copy case is given by
\begin{equation}
\begin{split}
R_0(q_1)=&p_0^1p_{0|0}^{2,1}R_2(q_{2|00})+p_0^1p_{1|0}^{2,1}R_2(q_{2|01})\\&+p_1^1p_{0|1}^{2,1}R_2(q_{2|10})+p_1^1p_{1|1}^{2,1}R_2(q_{2|11})\;,
\end{split}
\end{equation}
where $p_i^1$ is the probability of the POVM $\Pi_i$ clicking, $p_{i|j}^{2,1}$ is the probability of the POVM $\Pi_i$ clicking in the second measurement given that the POVM $\Pi_j$ clicked in the first measurement and $q_{2|ij}$ is the posterior probability when $\Pi_i$ clicked in the first measurement and $\Pi_j$ clicked in the second measurement. By numerically searching over all angles we can verify the correctness of our results in the two-copy case. However, even in the three-copy case a numerical search is already very slow compared to dynamic programming.

Finally, we note that in Figs.~\ref{fig:delta},~\ref{fig:data} and \ref{fig:apend:exp} we can observe points where the gradient of the LOCC error probability is not continuous. It is worth noting that the discrete nature of Eq.~\eqref{eq:min} is responsible for this behaviour.

\section{Gap between LOCC measurements and collective measurements}
\label{apen:delta}
In Fig.~\ref{fig:delta} b), we plotted the difference between the expected error probability when using LOCC measurements and collective measurements, $\Delta_M$ in Eq.~\eqref{eq:delta}. In Fig.~\ref{fig:apendelta} we plot $\Delta_2$ for several different $v$ values.


\section{Further Experimental Results}
\label{apen:extres}
\begin{figure*}[t!]
\includegraphics[width=0.8\textwidth]{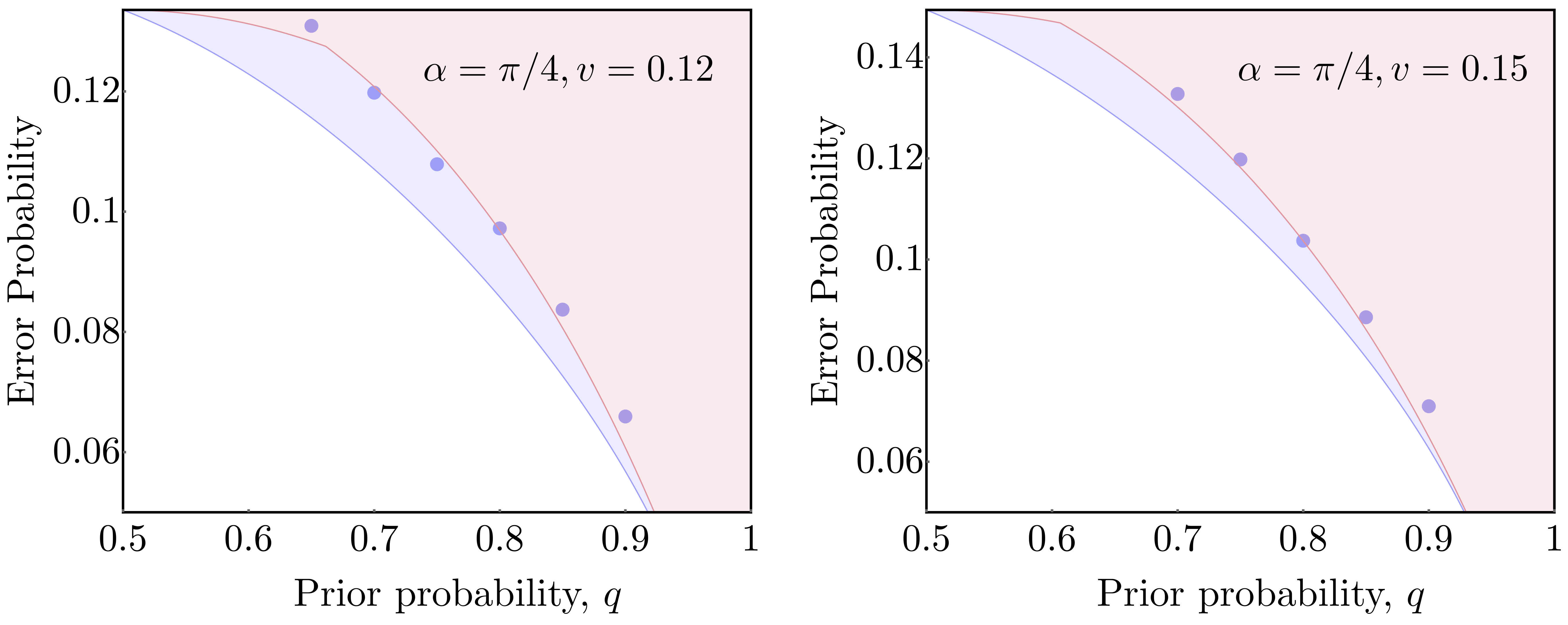}
\caption{\textbf{Further experimental data for discriminating quantum states using two-copy collective measurements.} As in Fig.~\ref{fig:data}, the shaded red and blue regions correspond to the limits of LOCC and collective measurements respectively. The blue markers correspond to the experimental data. Each data point corresponds to 200,000 shots. All data points included statistical error bars corresponding to one standard deviation obtained via bootstrapping, however in most cases the data points are larger than the error bars.}
\label{fig:apend:exp}
\end{figure*}

In Fig.~\ref{fig:apend:exp}, we show experimental results for $\alpha=\pi/4, v=0.12$ and $\alpha=\pi/4, v=0.15$. These results are worse than the results in the main text and the collective measurement performs close to the LOCC limit. There are several possible reasons for the worse performance with this set of parameters. It may be that the noise $v$ is introduced in an imperfect manner, so that experiments with larger $v$ values display worse performance. Additionally, as $v$ changes the gap between the optimal collective measurement and the optimal LOCC measurement changes, which may make it harder to observe any advantage of the collective measurement.

\bibliography{state_disc_bib}
\bibliographystyle{naturemag}

\end{document}